\documentclass[journal=jpccck,manuscript=article,layout=twocolumn]{achemso}

\usepackage[T1]{fontenc}
\usepackage{graphicx}
\usepackage{amssymb}
\usepackage{hyperref}
\usepackage{natbib}

\title{Tunneling-induced translation of intact $\pi$-radical clusters on Au(111)}

\author{Jacob D. Teeter}
\email{jacob.teeter@jku.at}
\author{Kateryna Averchenko}
\author{Maximilian Eliasch}
\author{Stefan M\"{u}llegger}
\affiliation{Solid State Physics Division, Johannes Kepler University Linz, 4040 Linz, Austria.}

\begin{document} 

\newcommand{\didv}{$\mathrm{d}I/\mathrm{d}V$\,}

\begin{abstract}
The scanning tunneling microscope (STM) is a powerful tool for investigating and manipulating molecules on surfaces. 
We demonstrate with a low-temperature STM operated at 6.2~K the controlled manipulation of ternary clusters of persistent molecular $\pi$ radicals as a whole. 
The ternary clusters -- each self-assembled from three $\alpha,\gamma$-bisdiphenylene-$\beta$-phenylallyl (BDPA) molecules on Au(111) -- maintain their natural cluster structure throughout tip-induced translation and rotation relative to the surface. 
Sustained and repeated dragging of radical clusters is shown to facilitate the construction of artificial assemblies of several clusters. 
Our results provide new opportunities for the creation and investigation of radical-based spin assemblies on surfaces.
\end{abstract}

\maketitle

\section{Introduction}
Scanning tunneling microscopy (STM) manipulation has enabled the well-controlled lateral translation of individual atoms \cite{Eigler1991} and molecules \cite{Bartels1997} on single-crystal metal surfaces. \cite{Hla2005,Ko2019,Moresco2004}
Organic molecules with persistent (stable) $\pi$ radicals bound to a surface \cite{MasTorrent2012} are particularly interesting from a scientific point of view, e.g. as model systems for low-dimensional spin ensembles \cite{Varela2023,Cho2024,Elvira2025,Trainer2022} or pure organic magnets \cite{Bocquet2019}. 
Thus, as a next step, the well-controlled formation of artificial radical clusters appears particularly appealing. 
First successful showcases of lateral STM manipulation of individual $\pi$-radicals have enabled the controlled build-up of artificial radical clusters in a one-by-one manner. \cite{Calmettes2014,Amokrane2017} 
Even the controlled lateral movement of weakly interacting supramolecular assemblies of (a few) noncovalently bonded molecules has been demonstrated without losing the collective integrity of the individual clusters. \cite{Nickel2013}
However, the controlled STM manipulation of intact radical clusters still has remained elusive. 

Here we show that supramolecular clusters of stable $\pi$-radicals translate as intact whole upon lateral STM manipulation, preserving their native structures. 
With the help of a low-temperature STM, we demonstrate the controlled positioning of radical clusters on Au(111) for building artificial lattices of radical clusters -- with potential implications for the way artificial spin structures are created on surfaces. \cite{Bocquet2019,Xu2020}

\section{Results}
\begin{figure}
\includegraphics[width=8.6cm]{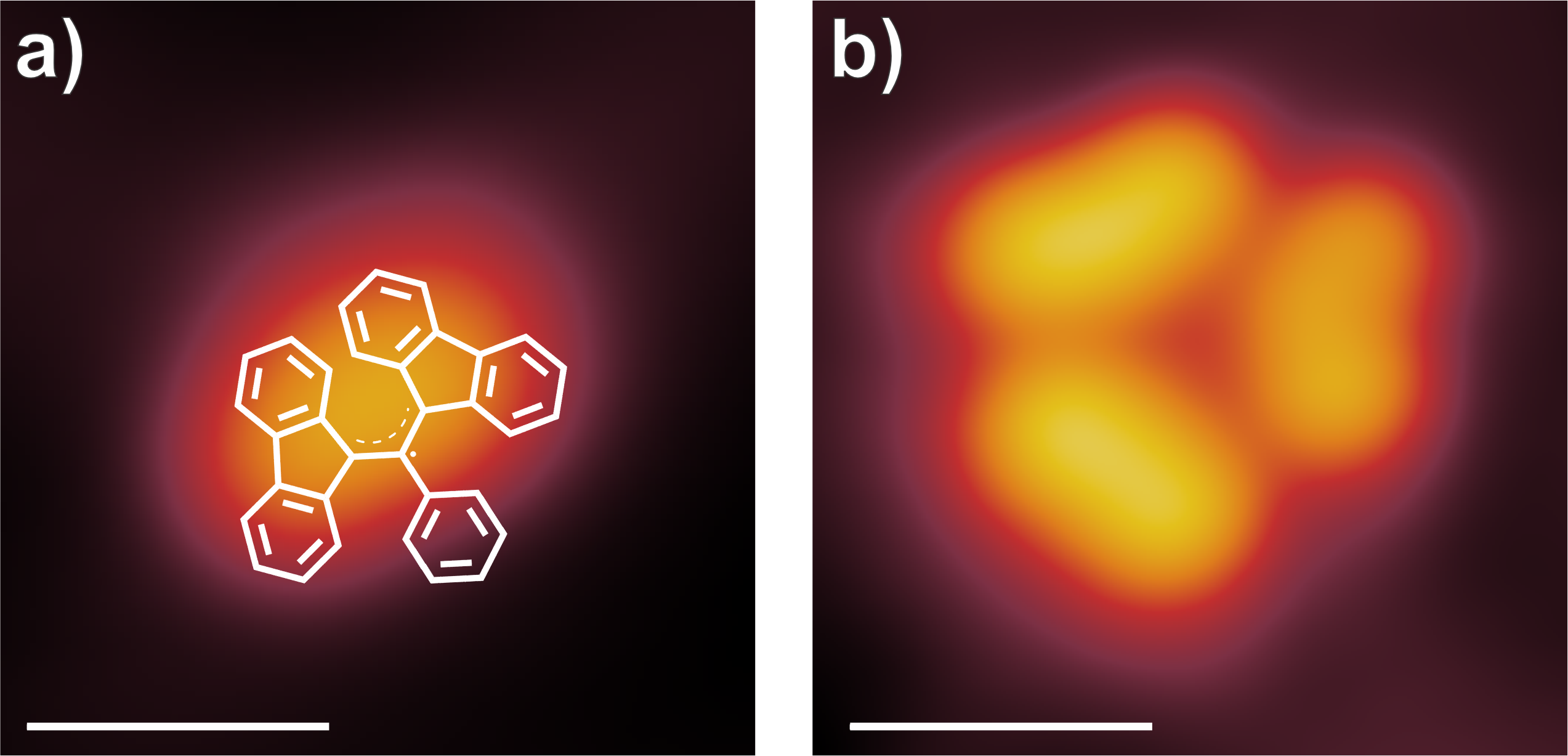}
\caption{\label{fig:tern} (a) Single BDPA molecule adsorbed on Au(111) as seen in STM in constant-current mode ($-1$~V, $150$~pA) with molecular structure superimposed to scale. (b) Ternary cluster of BDPA molecules on Au(111) as seen in STM in constant-current mode ($-0.01$~V, $500$~pA). $z$-ranges are $0-200$~pm and scale bars are 1~nm. Acquisition temperature 6.2~K.} 
\end{figure}

As model system we have chosen BDPA ($\alpha,\gamma$-bisdiphenylene-$\beta$-phenylallyl, $\mathrm{C_{33}H_{21}}$) -- also known as the Koelsch radical \cite{Koelsch1957}. 
The individual BDPA molecule maintains its unpaired $\pi$ electron state upon adsorption on Au(111) \cite{Mullegger2013a} as well as in a single-molecule-junction configuration \cite{Wang2025}.
Fig.~\ref{fig:tern}(a) shows a single BDPA molecule adsorbed over an fcc region of the $22\times\sqrt{3}$ reconstructed \cite{Barth1990} Au(111) surface as imaged by STM at 6.2~K -- with the molecule's chemical structure overlaid to scale as a guide to the eye. 
Depositing 0.03~monolayers of BDPA on Au(111) at room temperature causes individual BDPA molecules to preferentially form ternary clusters as reported earlier by our group. \cite{Mullegger2012b} 
At such low coverages, ternary clusters prefer to nucleate over fcc regions. 
Fig.~\ref{fig:tern}(b) shows a characteristic example of such a ternary cluster on fcc Au(111) as seen by STM at 6.2~K. 
We find stable tunnel conditions for imaging ternary clusters within a voltage range of $|V| \le 1.5$~V and up to 650~pA at constant-current conditions. 

\subsection{Manipulation of ternary clusters}
\begin{figure}
\includegraphics[width=8.6cm]{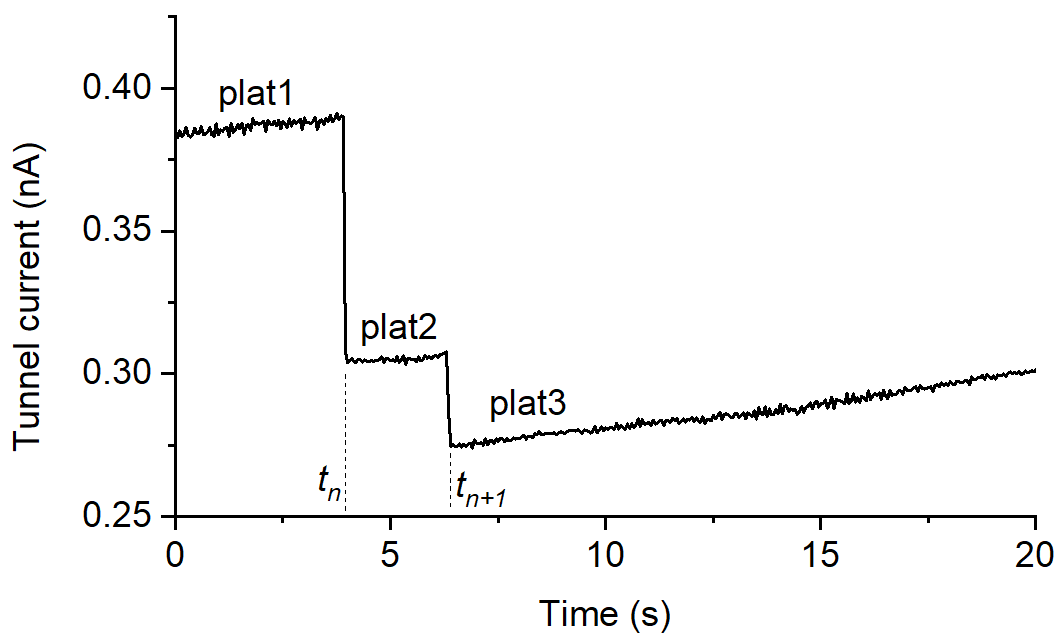}  
\caption{\label{fig:trace1} Time dependence of tunnel current, $I(t)$, after deactivating the feedback and applying a tunnel voltage pulse of $+2$~V (see text) with the STM tip positioned over the center of a ternary cluster of BDPA on Au(111). Discrete jumps in $I(t)$ occur at specific times labeled.} 
\end{figure}

After deactivating the feedback loop at imaging conditions, we operate the STM in constant-height mode and position the tip at a fixed lateral position $(x,y)$ over a ternary cluster. 
Increasing the tunnel voltage above a threshold of $|V|\geq2$~V (voltage pulse as described in methods section) results in the tunnel current $I(t)$ becoming unsteady; in particular, starting to exhibit discrete jumps as shown in Fig.~\ref{fig:trace1}. 
Similar jumps observed by STM on weakly surface-adsorbed molecules have been interpreted as signs of translational \cite{Bartels1997} and/or rotational \cite{Simpson2019} movement of the molecule relative to the substrate atomic lattice \cite{Simpson2023}. 

\begin{figure}
\includegraphics[width=8.6cm]{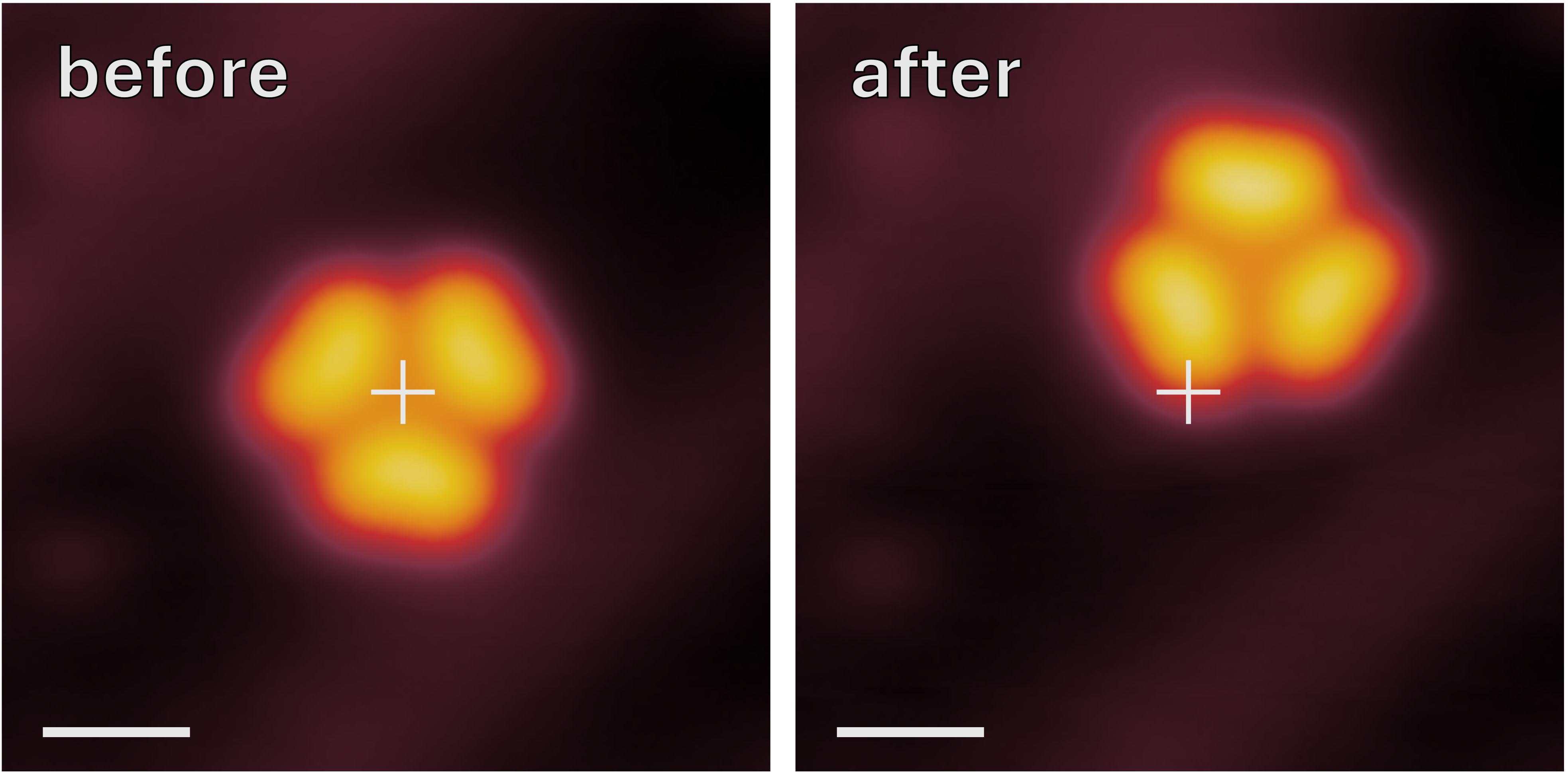}
\caption{\label{fig:manip0} Ternary cluster of BDPA on Au(111) as seen by STM at 6.2~K ($-0.8$~V, $150$~pA, $z$-scale $0-200$~pm and scale bars are 1~nm) before (left) and after (right) undergoing lateral translation and rotation relative to the Au(111) surface induced by a voltage pulse of $+2$~V for 20~s with deactivated feedback (see methods) causing the discrete jumps in the $I(t)$ curve shown in Fig.~\ref{fig:trace1}. The $+$ marks the tip position during voltage pulse.} 
\end{figure}

Imaging a ternary cluster before and after observing a discrete jump in the $I(t)$ trace induced by a voltage pulse reveals that the cluster has indeed undergone lateral  translational and/or rotational movement relative to the Au(111) surface.
Fig.~\ref{fig:manip0} shows a representative showcase of such a roto-translation as seen by STM. 
Based on our experimental STM data, we attribute the individual steps in $I(t)$ to translations and rotations of the cluster relative to the Au(111) surface; in addition, we rule out other possible origins for the jumps, such as, formation of a single molecule junction \cite{Haiss2004,Aragones2016} and instability of the STM tip \cite{Huang1995}. 

Intriguingly, and clearly evidenced by our STM data (Fig.~\ref{fig:manip0}), the ternary cluster moves as a whole upon voltage pulse manipulation, i.e. all three individual molecules of a cluster move in a concerted manner and maintain the intact structure of the ternary cluster after manipulation -- rather than movement of individual molecules separately. 
In our experiments, we never observed a separation of splitting-off of individual molecules off a cluster and, furthermore, detailed analysis yields no significant change in topographic height or preferred orientation of the clusters after undergoing (roto)translation.

\subsection*{Clusters translate towards the tip initially being positioned outside}
\begin{figure}
\includegraphics[width=8.6cm]{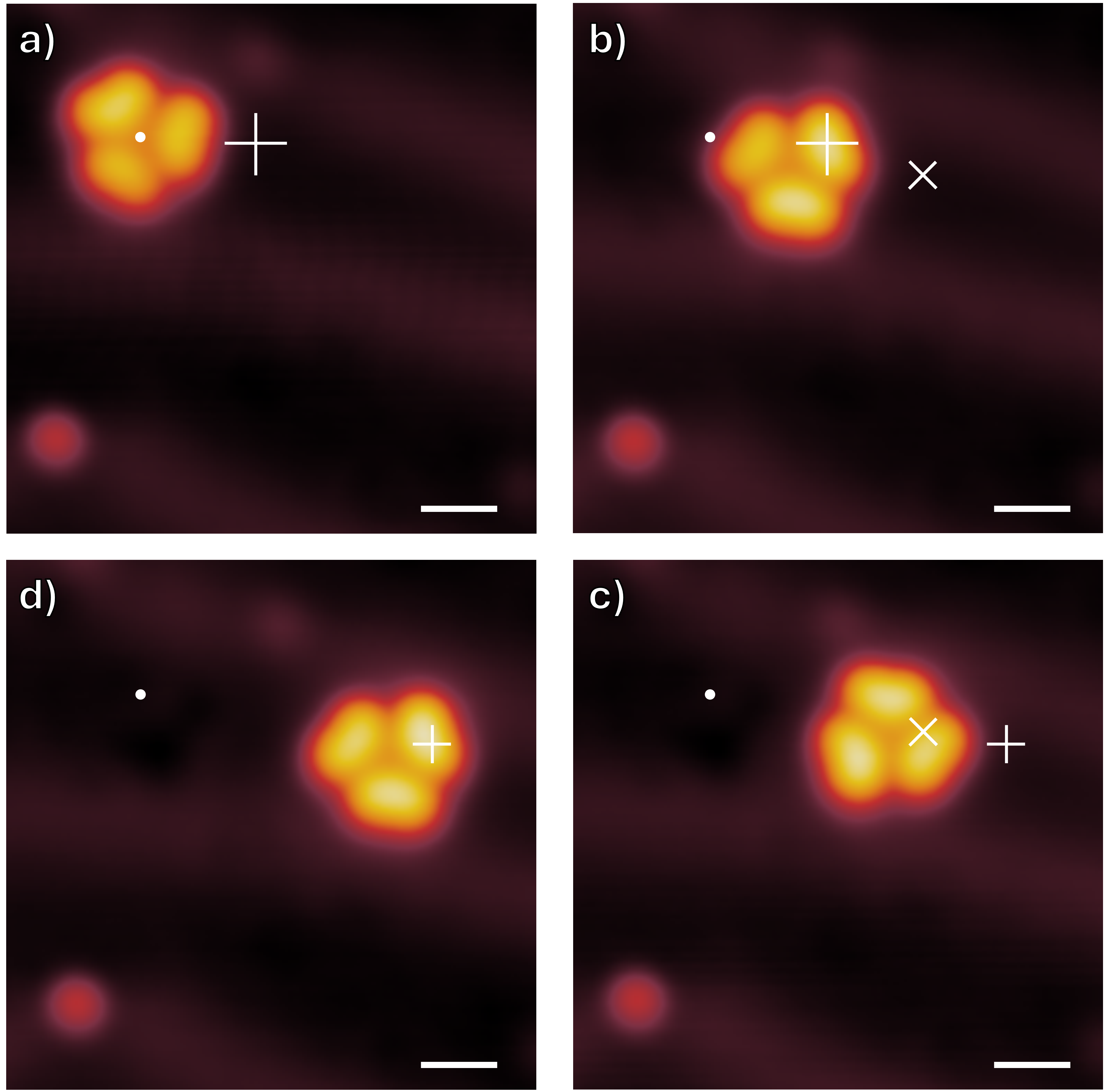}
\caption{\label{fig:follow} STM images of sequential roto-translation of a ternary cluster, wherein the cluster is displaced in the direction toward the STM tip position. The small white dot marks the center of the ternary cluster in its starting position. The white $+$ in (a) marks the tip location during a voltage pulse performed adjacent to the ternary cluster, and the mark is included again in (b) as a guide to the eye. The white $\times$ in (b) marks the location of a second voltage pulse, and it is included again in (c) as a guide to the eye. A final pulse was performed at the location marked in (c) with a small white $+$, and it is included again in (d) for reference. All pulses were at $+2$~V for 2~s, performed with the feedback loop opened at $-50$~mV and 650~pA. Scan parameters: $-0.8$~V, $650$~pA, $z$-scale is $0-200$~pm. All scale bars are 1~nm. } 
\end{figure}

In addition to manipulation induced by pulses applied directly over the center-of-gravity of a ternary cluster, we also demonstrate that ternary clusters translate towards the location of the tip if a pulse is performed within a distance of $\approx1.5$~nm from the center.
Fig.~\ref{fig:follow} shows a sequence of STM images of this type of pulse-driven manipulation of a ternary BDPA cluster. 
The tip was positioned at the location marked in (a) with a $+$ at a setpoint of $-50$~mV and 650~pA, and then a 2~s pulse at $+2$~V was applied. 
After re-imaging the very same image frame, the ternary cluster was observed to have translated towards the tip position during the pulse (see white dot in all four panels as reference for the initial location of the cluster). 
The cluster was translated further to the right in (b) and (c) with subsequent pulses applied through the tip at the locations marked with a white $\times$ and a second $+$ symbol, respectively.

An attractive force towards the STM tip was previously explained by the electric force between a nonzero molecular electric dipole and the electric field near the STM tip. \cite{Kuehne2020,Simpson2023}
In the case of the ternary clusters, we rationalize the observed attractive force towards the STM tip by similar arguments. 
The strong inhomogeneous electric field -- reaching the order of $10^{9}$~V/m near the center of the tunnel junction during the voltage pulse -- induces in the BDPA molecules of the cluster an electric dipole moment $\vec{p}$ with a nonzero lateral component. 
This results in a net electric force\cite{Griffith_Edyn} $\vec{F} = (\vec{p} \cdot \vec{\nabla})\cdot \vec{E}$ on the molecule that is directed along the electric field gradient towards the STM tip. 
Notice that the intermolecular attraction between the molecules of the cluster is stronger than the electric force, allowing the cluster to move as a whole -- as observed by STM. 
This points to a substantially larger intermolecular (attractive) interaction between the molecules within the cluster as compared to the surface diffusion barrier -- the latter being effectively lowered by the electric field near the tunnel junction. \cite{Kuehne2020}

\subsection*{Controlled building of artificial strucutre}
\begin{figure}
\includegraphics[width=8.6cm]{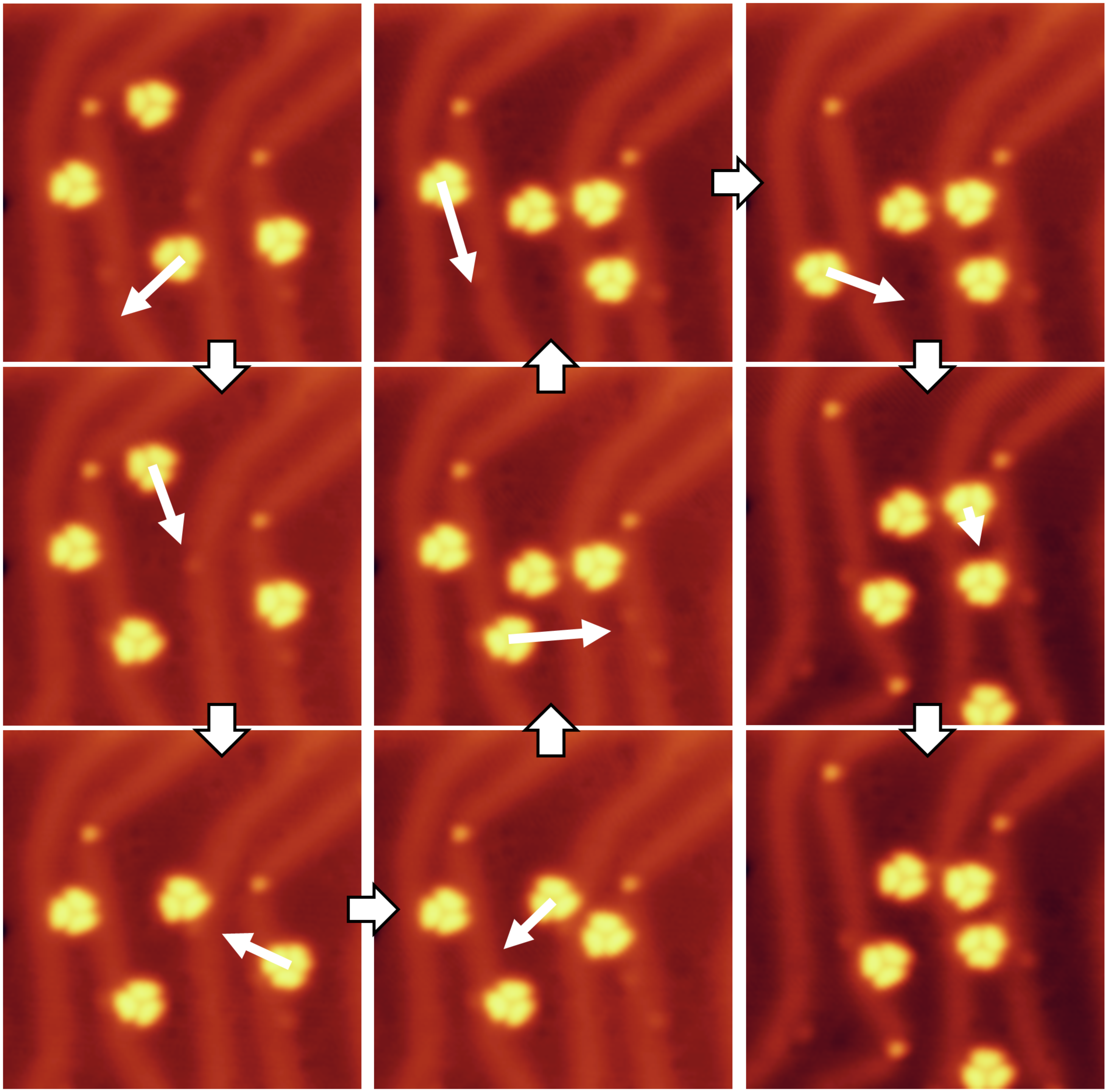}
\caption{\label{fig:drag} Sequence of STM images showing the controlled sustained and repeated dragging of ternary clusters of BDPA on Au(111) at 6.2~K (see text).} 
\end{figure}

Based on the success of pulse-based manipulation of intact ternary clusters, and operating on the same principle, the next logical step was to attempt sustained dragging with the STM tip. 
With the STM tip positioned near the edge of a cluster at $+0.8$~V and 650~pA, the feedback loop was disengaged, and the voltage was set to $+2$~V. 
Increased tip-cluster interaction is indicated by a marked increase in the tunnel current to about 8 to 10~nA -- sufficient for dragging. 
After this, with a lateral tip movement speed of $\le 1.5$~nm/s, it was possible to drag clusters over a scale of several nanometers without damaging them or dropping them. 
Returning the bias voltage to 0.8~V results in the current returning to a lower value, and imaging afterwards reveals the result of the dragging step. 
Fig.~\ref{fig:drag} demonstrates a sequence of such dragging events where a cluster can be manipulated several times and in various different directions. 
Notably dragging has also been achieved over the $\approx0.2$~\AA~ high ridges \cite{Barth1990} of the $22\times\sqrt{3}$-reconstructed Au(111) herringbone reconstruction. 
In this fashion, we have succeeded in constructing artificial assemblies of ternary clusters that do not occur naturally through self-assembly.

\section{Conclusion}
In conclusion, we have demonstrated the well-controlled lateral manipulation (translation) by STM of ternary clusters of stable $\pi$ radicals on Au(111) -- while fully maintaining their native structure. 
This facilitates the controlled building of artificial nanostructures out of individual ternary clusters. 
We foresee impact of our results for the future rational design of artificial lattices of radicals on surfaces. 

\section{Materials and Methods} 

BDPA was obtained from Sigma-Aldrich as a 1:1 complex with benzene, and was degassed under high vacuum conditions at 75°C overnight prior to deposition.
BDPA was deposited from a home-built crucible with a temperature of circa 96°C onto room-temperature Au(111) at a rate of 0.03 monolayers per minute. 
The substrate employed was an Au(111) single crystal (accuracy $<0.1$°; surface roughness of $<0.01$~$\mu$m). 
It was prepared by repeated Ar$^{+}$ ion sputtering and annealing to 600°C. 
Ultra-high vacuum scanning tunneling microscopy was performed using a commercial Omicron Polar STM at 6.2~K. 
The STM tip was a homemade electrochemically etched tungsten wire which was annealed after introduction to vacuum in order to remove oxide contamination. 
Bias voltage was applied to the sample. 
Tip cleanliness was ensured via repeated sub-nanometer indentation into the Au(111) surface.

For inducing lateral manipulation of ternary clusters, we start from well-defined initial conditions and proceed as follows: 
(1) safe imaging setpoint of 650~pA, $<|1.5|$~V.
(2) Set $I=650$~pA and $V=-50$~mV. 
(3) Position the STM tip over the center-of-mass of a ternary cluster. 
(4) Deactivate the feedback loop. 
(5) Apply voltage pulse of $+2$~V for several seconds. 
The probability of displacement using the above manipulation conditions, with the tip positioned either directly above or $\approx1.5$~nm off the center of the cluster, is found to be $\geq85$\%.

\section{Author contributions}
J.D.T. acquired collected data, performed data analysis, and contributed to the original draft; 
K.A. contributed to original drafting; 
M.E. performed data analysis; 
S.M. supervised the project, acquired funding, and contributed to original draft. 
All authors contributed to the final revisions and approved the submitted version.

\section{Acknowledgement}
The authors acknowledge the financial support of the European Research Council (ERC Consolidator Grant 771193) and the Government of the Province of Upper Austria (WI-2018-539817) together with the Johannes Kepler University Linz (LIT-2016-1-ADV-002).


\providecommand{\latin}[1]{#1}
\makeatletter
\providecommand{\doi}
  {\begingroup\let\do\@makeother\dospecials
  \catcode`\{=1 \catcode`\}=2 \doi@aux}
\providecommand{\doi@aux}[1]{\endgroup\texttt{#1}}
\makeatother
\providecommand*\mcitethebibliography{\thebibliography}
\csname @ifundefined\endcsname{endmcitethebibliography}
  {\let\endmcitethebibliography\endthebibliography}{}

\end{document}